\documentclass{elsart3p}
\usepackage{amssymb}
\usepackage{graphicx}
\begin{document}
 \begin{frontmatter}


 
\title{Pattern recognition and PID for COMPASS RICH-1}

 
\author[saclay]{P.Abbon},
\author[alessandria]{M.Alexeev\thanksref{leave-jinr}},
\author[munich]{H.Angerer},
\author[infn-trieste]{R.Birsa},
\author[lip]{P.Bordalo\thanksref{also-ist}},
\author[trieste]{F.Bradamante},
\author[trieste]{A.Bressan},
\author[torino]{M.Chiosso},
\author[trieste]{P.Ciliberti},
\author[infn-torino]{M.L.Colantoni},
\author[saclay]{T.Dafni},
\author[infn-trieste]{S.Dalla Torre},
\author[saclay]{E.Delagnes},
\author[infn-torino]{O.Denisov},
\author[saclay]{H.Deschamps},
\author[infn-trieste]{V.Diaz},
\author[torino]{N.Dibiase},
\author[trieste]{V.Duic},
\author[erlangen]{W.Eyrich}, 
\author[torino]{A.Ferrero},
\author[prague]{M.Finger},
\author[prague]{M.Finger Jr},
\author[freiburg]{H.Fischer},
\author[munich]{S.Gerassimov},
\author[trieste]{M.Giorgi},
\author[infn-trieste]{B.Gobbo},
\author[freiburg]{R.Hagemann},
\author[mainz]{D.von~Harrach},
\author[freiburg]{F.H.Heinsius},
\author[bonn]{R. Joosten},
\author[munich]{B.Ketzer},
\author[cern]{V.N. Kolosov\thanksref{leave-ihep}},
\author[freiburg]{K.K\"onigsmann}, 
\author[munich]{I.Konorov},
\author[liberec]{D.Kramer},
\author[saclay]{F.Kunne},
\author[erlangen]{A.Lehmann},
\author[trieste]{S.Levorato},
\author[infn-torino]{A.Maggiora},
\author[saclay]{A.Magnon},
\author[munich]{A.Mann},
\author[trieste]{A.Martin},
\author[infn-trieste]{G.Menon},
\author[freiburg]{A.Mutter},
\author[bonn]{O.N\"ahle},
\author[freiburg]{F.Nerling}, 
\author[saclay]{D.Neyret},
\author[alessandria]{D.Panzieri},
\author[munich]{S.Paul},
\author[trieste]{G.Pesaro},
\author[erlangen]{C.Pizzolotto},
\author[liberec,infn-trieste]{J.Polak},
\author[saclay]{P.Rebourgeard},
\author[saclay]{F.Robinet},
\author[torino]{E.Rocco},
\author[trieste]{P.Schiavon}, 
\author[freiburg]{C.Schill},
\author[erlangen]{P.Schoenmeier},
\author[erlangen]{W.Schr\"oder},
\author[lip]{L.Silva},
\author[prague]{M.Slunecka},
\author[trieste]{F.Sozzi\corauthref{cor}},
\corauth[cor]{Corresponding author: federica.sozzi@ts.infn.it}
\author[prague]{L.Steiger},
\author[liberec]{M.Sulc},
\author[liberec]{M.Svec},
\author[trieste]{S.Takekawa},
\author[infn-trieste]{F.Tessarotto},
\author[erlangen]{A.Teufel},
\author[freiburg]{H.Wollny}

\address[alessandria]{INFN, Sezione di Torino and University of East Piemonte, Alessandria, Italy}
\address[bonn]{Universit\"at Bonn, Helmholtz-Institut f\"ur Strahlen- und Kernphysik, Bonn, Germany}
\address[cern]{CERN, European Organization for Nuclear Research, Geneva, Switzerland}
\address[erlangen]{Universit\"at Erlangen-N\"urnberg, Physikalisches Institut, Erlangen, Germany}
\address[freiburg]{Universit\"at Freiburg, Physikalisches Institut, Freiburg, Germany}
\address[liberec]{Technical University of Liberec, Liberec, Czech Republic}
\address[lip]{LIP, Lisbon, Portugal}
\address[mainz]{Universit\"at Mainz, Institut f\"ur Kernphysik, Mainz, Germany}
\address[munich]{Technische Universit\"at M\"unchen, Physik Department, Garching, Germany}
\address[prague]{Charles University, Praga, Czech Republic  and JINR, Dubna, Russia}
\address[saclay]{CEA Saclay, DSM/DAPNIA, Gif-sur-Yvette, France}
\address[torino]{INFN, Sezione di Torino and University of Torino, Torino, Italy}
\address[infn-torino]{INFN, Sezione di Torino, Torino, Italy}
\address[trieste]{INFN, Sezione di Trieste and University of Trieste, Trieste, Italy}
\address[infn-trieste]{INFN, Sezione di Trieste, Trieste, Italy}
\thanks[leave-jinr]{on leave  from JINR, Dubna, Russia}
\thanks[also-ist]{also at IST, Universidade T\'ecnica de Lisboa, Lisbon, Portugal}
\thanks[leave-ihep]{on leave  from IHEP, Protvino, Russia}

\begin{abstract}
  A package for pattern recognition and PID by COMPASS RICH-1 has been developed and used for the analysis
  of COMPASS data collected in the years 2002 to 2004, and 2006-2007 with the upgraded RICH-1 photon detectors. 
  It has
  allowed the full characterization of the detector in the starting version and 
  in the upgraded one, as well as the PID for
  physics results.
  We report about the package structure and algorithms, and the detector characterization and PID results.
% Text of abstract
\end{abstract}

\begin{keyword}
COMPASS \sep RICH \sep multi-anode
photomultiplier \sep particle identification \sep reconstruction algorithms
\PACS 29.40.Ka \sep 42.79.Pw \sep 07.05.Kf \sep 29.85.-c
\end{keyword}
\end{frontmatter}

% main text
\section{Introduction}

COMPASS~\cite{compass} 
is a fixed target experiment at CERN SPS.
The COMPASS  physics program is focused mainly on the study of the nucleon spin structure and on 
  hadron spectroscopy.
COMPASS has collected data in   2002-2004 and 2006-2007 using a muon beam of 160~GeV/c. 
The apparatus includes a polarized target for spin structure studies, 
a large acceptance double spectrometer, electromagnetic and 
hadron calorimeters and muon filters \cite{Abbon:2007pq}.
Charged hadron identification (PID) is performed by means of 
a focusing Ring Imaging CHerenkov detector of large dimensions, the COMPASS RICH-1~\cite{Albrecht:05}.
RICH-1 is a Cherenkov imaging detector with large angular acceptance 
($\pm$180~mrad vertical, $\pm$250 mrad horizontal).
The radiator gas is C$_4$F$_{10}$ at atmospheric pressure and the typical particle path
length in the radiator is 3~m. Image focusing is obtained by 2 spherical surfaces, 
formed by hexagonal and pentagonal mirror elements, 
resulting in a reflecting surface with a total area larger than 20~m$^2$.
Images are collected on two sets of photon detectors, placed above 
and below the detector acceptance region.
Until 2004, the photon detectors used were eight MWPCs with segmented CsI photo-cathodes, covering 
a surface of 5.2~m$^2$. 
In 2006, an important  upgrade of the photon detectors  system
was completed. The upgrade  is based on two complementary techniques  for the central and peripheral 
regions of the photon detectors.
The central part, corresponding to four of the 16 MWPC photo-cathodes,
has been equipped with multi-anode photomultiplier (MAPMT)  with UV 
extended glass window (Hamamatsu R7600-03-M16) coupled to individual telescopes of fused silica lenses
and a fast digital read-out system~\cite{Abbon:2006bb};
in the peripheral part, a new readout system~\cite{APV-rich},
based on the APV chip~\cite{APV}, almost dead-time free and characterized by 
improved time resolution, has been installed on the
already existing CsI MWPCs.

 From the   point of view of the software data handling, 
the upgrade results in two main consequences: 
\begin{itemize}
\item The geometry and the spatial resolution of the 
  central and peripheral detectors are different: the size of the pads in the 
peripheral part is 8$\times$8~mm$^2$, while in the central part the {\it pseudopads} (the MAPMT pixels
projected through the optical telescope onto the plane previously 
housing the CsI photo-cathodes)  have a dimension of about 12$\times$12~mm$^2$;
the  different spatial resolution is one of the elements that requires  
different detector characterization in the two parts.
\item The spectrum of the detected Cherenkov photons is different in the central part (200-750~nm)
respect to the spectrum of the peripheral part (165-200~nm); this implies that the 
average effective value of the radiator refractive index
is different according to the type of the photon detector involved; the mean values
of the effective refractive index, as
 evaluated offline from a data sample,   are respectively:
$n-1\sim0.001345$ for the MAPMT detectors and $n-1\sim0.001528$ for the CsI MWPC detectors.
The difference in the effective refractive indexes implies also that the rings
detected in   the MAPMT and the CsI region,    have different radii: the radius of the 
rings detected in the peripheral region are about 30\% larger  than those detected in the MAPMT
region.
 \end{itemize}

\section{RICH-1 software}
The analysis of the RICH data can be divided in two main steps: the ring reconstruction and the
particle identification. The aim  of the RICH in the experiment  is to give an answer
to the latter point, nevertheless a complete understanding of the RICH performances is needed
before any attempt to identify particles, and for this purpose 
the ring reconstruction is largely used.
 
The reconstruction of the RICH-1 data is performed with RICHONE, a dedicated class 
 of the COMPASS reconstruction programme, CORAL. 
As starting point of the reconstruction, 
 the RICH digits in both the MAPMT and the CsI MWPC part are used;  
the hits  are rearranged in clusters in the 
CsI MWPC part only, since the MAPMT 
crosstalk is completely negligible.
Using the cluster position on the detector surface, 
and  assuming as emission point the 
middle point of the particle trajectory inside the radiator,
the photon emission angles respect to the
particle trajectory $(\theta_{Ch}, \phi_{Ch})$ are computed using the
Hough transformation, following a recipe well 
known in the literature~\cite{YS}.
The trajectory is provided by the tracking package in the CORAL 
reconstruction programme.

The ring reconstruction is mainly used for the analysis of the RICH performances.
For all the clusters contained in a  region
of 70~mrad around the particle trajectory, the emission angles $(\theta_{Ch}, \phi_{Ch})$
are computed. 
Since a ring is characterized by a fixed value of  $\theta_{Ch}$,  
a peak in the $\theta_{Ch}$ distribution is searched for
through a scan with a fixed size window of $\pm3 \sigma_{ph}$, 
where $ \sigma_{ph}$ is the single photon resolution.
For the ring reconstruction, the Cherenkov angles of the photons detected in the MAPMT region
are normalized to the  angles in the CsI MWPC region, so as to combine, for rings not completely
included in a single detector region,  
the information coming from the two different  regions.
One of the main advantages of the simple recipe described
above is the 
automatic handling of events of different complexity, as is the case of the split rings. In fact,
due to  the RICH-1 detector architecture, photons emitted by  particles scattered 
at small vertical angles, illuminate both mirror surfaces, 
and are then reflected partially onto the upper and partially onto the lower 
photon detector set, forming a full ring image in a detector set and a partial ring in the
other one.

 The particle identification algorithm relies on a likelihood  function built from all the
photons associated to the particle in the  fiducial region of 70~mrad.  
A Poissonian probability is used to describe the expected number of detected photons (both signal 
and background contribution) as a function of the parameters of the particle trajectory. 
The likelihood is computed for five mass hypothesis (${e},{\nu},{\pi},{K},{p}$)
and for the background, corresponding to the hypothesis  of absence of signal.
For a given mass M, and for N photons,
the likelihood expression reads:
  
{\small \begin{equation} 
  L_M = \frac{(S_M+B)^N}{N!}e^{-(S_M+B)} \prod_{j=1}^{N} \frac{s_M(\theta_j,\phi_j)+b(\theta_j,\phi_j)}{S_M+B} 
\end{equation} }
The background term, $b(\theta_j,\phi_j)$, is evaluated pad by pad from a map
of the integrated cluster distribution, taken from the data themselves.
$S_M$ and $B$ are the signal and background photons integrated on the particle fiducial region, defined above.
The signal term $s_M(\theta_j,\phi_j)$ is taken as a Gaussian, with mean value the Cherenkov angle 
evaluated from the kinematics corresponding to the mass M and with the width of the single photon
resolution:
\begin{equation}
s_M(\theta_j,\phi_j)=\frac{S_0}{\sqrt{2\pi}\sigma_{ph,\,j}}\cdot e^{-\frac{1}{2}\frac{(\theta_j-\Theta_M)^2}{\sigma^2_{ph,\,j}}}\epsilon(\theta_j,\phi_j);
\end{equation}
 $S_0$ is the expected number of photons evaluated using the Frank and Tamm equation; 
the $\epsilon$ term is the probability of detecting a photon, taking into account   the
dead zones in the detector.
 
\section{Detector characterization}
RICH-1  has been fully characterized in both the old and  the upgraded version, 
monitoring in operative conditions 
all the relevant observables: the number of detected photons, the angular resolution and  the PID efficiency.

\begin{figure}[!t]
\centering
\includegraphics[width=0.4\textwidth]{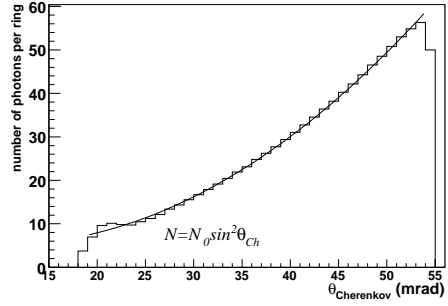}
\caption{\small{Number of photons per ring as a function of the Cherenkov angle, for rings 
detected in the MAPMT part only; the curve is a fit with a function of the type  
$N_0 \sin^2( \theta_{Ch})$. The number of photons at saturation  after background subtraction is around 56.  
\label{fig:np}} }
\end{figure}

The number of signal photons emitted at saturation has been evaluated through a fit 
of the number of   photons per ring   as a function of the 
Cherenkov angle, using 
a function of the type 
$N_0 \sin^2( \theta_{Ch})$. 
The number of detected photons at saturation is around 14 before the RICH-1 upgrade and, after the upgrade,
in the peripheral regions, 
while it has increased up to 
56(fig.~\ref{fig:np}) in the upgraded RICH, central regions.
Moreover, the ratio of the signal to the  background   in each  ring is increased   because  
the number of background clusters per ring has been reduced by a factor $\sim$30\% in the central part.

\begin{figure}[!t]
\centering
\includegraphics[width=0.4\textwidth]{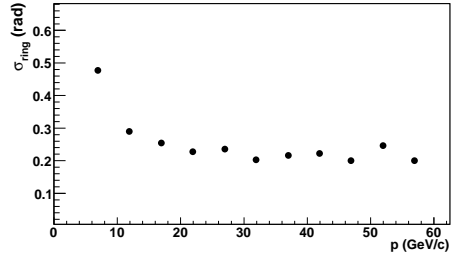}
\caption{\small{Standard deviation of the $\theta_{ring} - \theta_{\pi} $ 
distribution, for particles identified as pions, 
as a function of the particle momentum; ring detected in the MAPMT detector only.\label{fig:res}} }
\end{figure}
%\subsection{Angular resolution }
The single photon  resolution is evaluated from the width of the distribution
$\theta_{ph} - \theta_{\pi} $, where $\theta_{ph}$ is the Cherenkov angle of a photon belonging to 
the ring and $ \theta_{\pi} $ is the angle for the pion mass hypothesis.
In RICH-1 before the upgrade, the mean value of the resolution is around 1.2~mrad, 
while the resolution on the ring angle is around 0.6~mrad.
The two numbers do not scale with the square root of the number of detected photons per ring due to the large 
background contribution in each reconstructed ring,  diluting the signal.
After the upgrade, the single photon resolution in the central part is around 2~mrad,
while the ring angular resolution less than 0.3~mrad (fig.~\ref{fig:res}); 
the better scaling of the ring resolution with the number
of photons is due to the fact that the 
 background contribution per ring is small.
 
  \begin{figure}[!t]
   \centering
    \includegraphics[width=0.4\textwidth]{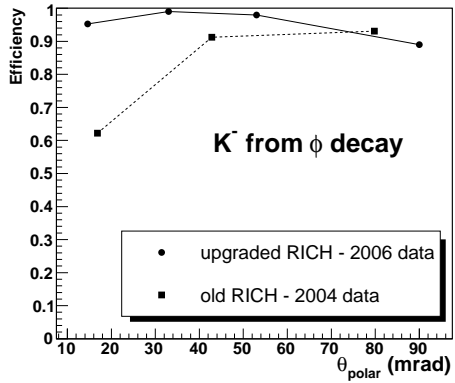}
    \caption{\small{ PID efficiency for K$^-$,  
       evaluated on a sample of particles 
       from $\phi$ decay, as a function of the particle polar angle. The two sets of data points 
       correspond to the old   and to the upgraded  versions of RICH-1.
       \label{fig:eff}} }
 \end{figure}
\begin{figure}[!t]
\centering
 \includegraphics[width=0.4\textwidth]{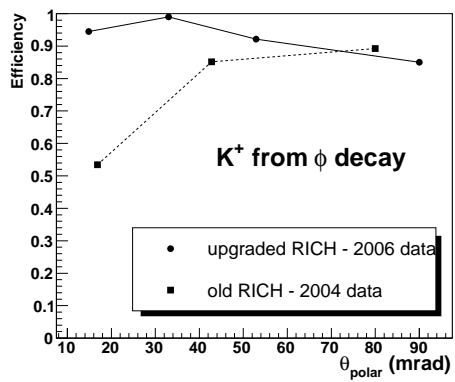}
\caption{Same as fig.~\ref{fig:eff}, for K$^{+}$ \label{fig:eff2}} 
\end{figure}
The PID efficiency has been evaluated selecting with kinematics criteria
a sample of exclusively produced $\phi$ mesons; the sample has a purity of about 90\%. 
The efficiency has been evaluated separately for $K^+$ and $K^-$,
using the positive or negative track coming from the $\phi$ decay.
In fig.~\ref{fig:eff} and~\ref{fig:eff2}, 
the  K efficiency is shown as a function of the particle polar angle, 
both for the old and for the upgraded RICH. The curve corresponding 
to the old RICH shows clearly that the efficiency increases at large
polar angle, since 
at small angles it is limited  
by the presence of an important background    coming from 
the muon beam halo. 
The impact of the upgrade on the RICH efficiency 
is clearly visible from the corresponding  efficiency curve:
also at small polar angle the efficiency is above 90\%.

\section{Conclusions}
The RICH-1   reconstruction software is based on a simple recipe, 
allowing an easy handling of events of different complexity.
The ring pattern recognition and the PID are fully independent and are used for different 
purposes: detector characterization and data analysis, respectively.
The complete RICH-1 characterization, both in the old and in the upgraded version,
has been presented and
confirms  a large improvement of the RICH-1 performance in its upgraded version. 
 
\section{Acknowledgments}
We acknowledge the support of BMBF (Germany) 
and of the European Community-Research 
Infrastructure Activity under the FP6
''Structuring the European 
Research Area'' programme 
(Hadron Physics, contract number RII3-CT-2004-506078).

\end{document}